\documentclass{article}
\begin{document}

\begin{center}
\centerline{\large \bf On the phase dependence of a reversed quantum 
transitions} 
 
\end{center}

\vspace{3 pt}
\centerline{\sl V.A. Kuz'menko\footnote{Electronic 
address: kuzmenko@triniti.ru}}

\vspace{5 pt}

\centerline{\small \it Troitsk Institute for Innovation and Fusion 
Research,}
\centerline{\small \it Troitsk, Moscow region, 142190, Russian 
Federation.}

\vspace{5 pt}

\begin{abstract}

Physical nature of widely known Ramsey fringes phenomenon is discussed.

\vspace{12 pt}
\end{abstract}

\vspace{8 pt}

Ramsey fringes are an oscillation in population transfer in a two-level 
quantum system under action of an external driving field, which is split 
in space or in time [1]. Although, the phenomenon is known for many years, 
it does not have clear physical explanation. However, such explanation is 
very simple: this is a result of natural property of a reversed quantum 
transition – its cross-section depends on the phase of driving field.

Let us discuss recent classical pump-probe experimental study of interaction 
of femtosecond laser pulses with Rb vapor [2]. When very weak collinear fs 
laser pulse interacts with unexcited atoms before the main pump pulse, the 
phase modulation is absent. But rather strong phase modulation exists when 
the probe pulse pass during or after the pump pulse and interacts also with 
the excited atoms. Such results usually are explained by a quantum 
interference of "wave-packets", which are formed by two laser pulses [3]. 
However, this is not a physical explanation. A photon has some energy. 
This energy can not disappear. When after the second laser pulse population 
of excited states vanish, it means that efficient stimulated emission takes 
place and we should say about high cross-section of reversed process. We 
have sufficient experimental proofs of strong inequality of differential 
cross-sections of forward and reversed transitions in optics [4].

The phase sensitivity is a property of only reversed transitions. The 
experiments clearly show that forward transition to any other quantum state 
does not have the specific dependence from the phase of the second laser 
pulse [5, 6]. Other beautiful experiment with diamond configuration of 
used quantum levels in Rb shows that forward and reversed pathway may be 
not coincide with each other [7]. Only the initial and final states of a 
quantum system must exactly coincide.  This demand can explain why we do 
not see efficient three photon mixing processes in a gas phase. Each photon 
has a spin. So, using odd number of photons it is impossible to return a 
quantum system exactly into the initial state. In a solid state a crystal 
lattice can eliminate the role of a spin and we can see efficient three 
photon mixing processes. In a gas phase efficient photon mixing processes 
are possible only with an even number of photons, that we can see, for 
example, in a high harmonic generation process [8]. 

Experimental study of reversed transitions is rather simple: in a common 
case we need in a pump-probe apparatus with a collinear sufficiently weak 
probe pulse, which does not considerably modify the population distribution 
of quantum states [9]. Unfortunately, such experimental results are 
practically absent in literature.

For a long time a scientists try to invent an empirical methods of so-called 
"coherent control" [10 - 13] instead of to study its physical base: 
inequality of forward and reversed processes or time noninvariance in 
quantum physics.

\end{document}